\documentclass[twocolumn,preprintnumbers,amsmath,amssymb]{revtex4}
\usepackage{graphicx}

\begin{document}

\title{Amplitude dependent frequency, desynchronization,  and stabilization in noisy metapopulation dynamics}

\author{Refael Abta$^{1}$, Marcelo Schiffer$^{2}$   and Nadav M. Shnerb$^{1}$}

\affiliation{ ($1$)  Department of Physics, Bar-Ilan University,
Ramat-Gan 52900 Israel  \\ ($2$) Department of Physics, Judea and
Samaria College, Ariel 44837 Israel.}

\begin{abstract}
The enigmatic stability of population oscillations  within
ecological systems has remained an open theoretical issue for over
seven decades. The key models for deterministic description of
population dynamics - the Lotka-Volterra prey-predator model and the
Nicholson-Bailey  host-parasitoid system - fail to support an
attractive manifold. Accordingly, in  any ecological system
described by these models, species must undergo extinction. Dozens
of studies regarding this ecological paradox have revealed the
important role played by spatial migration and noise, but have yet
to pinpoint the generic stabilizing process. This underlying
mechanism is presented and analyzed here in the framework of two
interacting species free to migrate between two spatial patches. We
show that the combined effects of migration and noise cannot account
for the stabilization. The missing ingredient is the dependence of
the oscillations' frequency upon their amplitude; with that,
noise-induced differences between patches are amplified due to the
frequency gradient. Migration among desynchronized regions then
stabilizes a "soft" limit cycle in the vicinity of the homogenous
manifold.  A simple model of diffusively coupled oscillators allows
the derivation of quantitative results, like the functional
dependence of the desynchronization upon diffusion strength and
frequency differences. Surprisingly, the oscillations' amplitude is
shown to be (almost) noise independent. The results are compared
with a numerical integration of the marginally stable Lotka-Volterra
equations. An unstable system is extinction-prone for small noise,
but stabilizes at larger noise intensity. The methodology presented
here may be applied to a wide class of spatially-extended ecological
problems. Indeed, this understanding will likely aid in designing
strategies for preventing spatial synchronization responsible for
species extinction.
\end{abstract}

\maketitle

The idea that species populations fluctuate in time has been well
known since the early days of history. Ancient day naturalists, like
Herodotus and Cicero,  perceived the persistence of prey species in
the face of adversity as a manifestation of divine power and the
creator's design \cite{cuddington}. In modern times, the
mathematical description of prey-predator interacting populations
was given, using deterministic, continuous time partial differential
equations, by Lotka and Volterra \cite{lotka,volterra,murray}, the
analogous model with discrete time step was introduced for a
parasitoid-host system by Nicholson and Bailey \cite{bailey}. Both
models allow, essentially, for population oscillations around a
steady state. As pointed out by Nicholson \cite{nic33}, these
oscillations are an intrinsic property of interacting populations.
If the density of the host, say, is above its steady value, it will
be reduced by the parasite. However, when the host reaches its
steady density, the density of parasites will be above its steady
value. "Consequently, there are more than sufficient parasites to
destroy the surplus hosts, so the host density is still further
reduced in the following generation ... Clearly, then, the densities
of the interacting animals should oscillate around their steady
value" \cite{nic33}. Oscillations in populations and metapopulations
have been observed in many field studies \cite{murray} and even in
controlled experiments \cite{kerr,holyoak}. The stabilization of
such oscillations is considered to be a major factor affecting
species conservation and ecological balance \cite{levin,blasius}.

Lotka, Volterra and Nicholson recognized  that the oscillations
described by their models are not stable \cite{nic33,cuddington}.
The Nicholson-Bailey map admits an unstable steady state where the
amplitude of oscillations grows exponentially with time; for the
Lotka-Volterra system, the fixed point is marginally stable,
rendering the system extinction-prone for any noise amplitude (See
Appendix 1). Indeed, experimental and theoretical studies of both
systems reveal that the oscillations increase in size until one of
the species becomes extinct \cite{gillaspee, gause, luckinbill}.

Nicholson \cite{nic33} was perhaps the first propose the  idea of
migration induced stabilization. Although on a single patch, the
oscillation amplitude grows in time and the system is driven to
extinction, desynchronization between weakly coupled spatial
patches, together with the effect of migration, leads to the
appearance of spatial patterns and stabilizes the global
populations. This seminal idea has been examined in many studies and
the main results, summarized in a recent review article
\cite{briggs}, are as follows:

\begin{itemize}

\item For any network of $N$ patches, if the migration between
patches is symmetric or almost symmetric (i.e., the diffusion of the
prey and the predator are, more or less, the same), there is no
diffusion induced instability, and the homogenous manifold is stable
\cite{allen,reeve,crowley}. Thus, the effect of migration alone does
not cure the instability problem. Diffusion induced instability may
occur if the migration rate of the predator is much smaller than
that of the prey \cite{jansen}, or in a case where the reaction
parameters vary on different spatial patches \cite{mordoch1,
mordoch2, hassel}.

\item Numerical simulations demonstrate, indeed, that Lotka-Volterra or
even Nicholson Bailey dynamics on spatial domains are much more
stable \cite{wilson, donaldson,agam,t1}. In general these numerical
experiments involve some sort of noise, like the intrinsic noise due
to the stochasticity associated with discrete individuals
(individual-based models), numerical noise etc.

\item  Accordingly, there is  broad agreement that  the
combined effect  of noise and diffusion is a necessary precondition
for population stabilization. However, up until now the qualitative
nature of the underlying mechanism has remained obscure, and no
theoretical framework that allows for quantitative prediction has
been presented.
\end{itemize}

This  theoretical gap may be addressed using a toy model for coupled
oscillators (see Appendix B). The main new ingredient emphasized by
the proposed model is the \emph{dependence of frequency on the
oscillation amplitude}, reflected by the gradient of the angular
velocity along the radius $\omega'(r)$. The instability induces
desynchronization iff the  small, noise-induced, differences between
patches are amplified  by the frequency gradient such that the
"desynchronization parameter" $\langle \phi^2 \rangle$ acquires a
finite value, leading to "restoring force" toward the origin of the
homogenous manifold.

The coupled oscillators model provides the basic theoretical
framework with  which  to explain the emergence of an attractive
manifold. In the context of the realistic, Lotka-Volterra model, the
lifetime of the system (time until extinction of one of the species)
is controlled by that manifold. A two-patch system is then described
by:
\begin{eqnarray}\label{two}
\frac{\partial a_1}{\partial t} &=& - \mu a_1 + \lambda_1  a_1 b_1 +
D_a (a_2-a_1) \nonumber \\   \frac{\partial a_2}{\partial t} &=& -
\mu a_2 + \lambda_1   a_2 b_2 + D_a (a_1-a_2)\\  \nonumber
\frac{\partial b_1}{\partial t} &=& \sigma b_1 - \lambda_2 a_1  b_1
+ D_b (b_2 - b_1) \\ \nonumber \frac{\partial b_2}{\partial t} &=&
\sigma b_2 - \lambda_2  a_2 b_2 + D_b (b_1 - b_2).
\end{eqnarray}
The invariant manifold is the two dimensional subspace $a_1=a_2, \ \
b_1 = b_2$. The time evolution of that system, with an  additive
noise, equal diffusivities, $D_a=D_b \equiv D$, and symmetric
reaction rates (See Appendix A) is obtained through Euler
integration. In the limit $D=0$ the patches are unconnected; thus,
starting from the homogenous fixed point $a_1 = a_2 = b_1 =b_2 =1$,
the single patch situation (Appendix A) still holds and the system
hits the absorbing walls after a characteristic, noise dependent,
time. In the opposite limit, $D=\infty$, the system sticks to the
invariant manifold and acts like a single patch (with modified noise
and interaction parameters), performing again  a random walk in the
invariant manifold. However, between these two extremes, there is a
region where the combined effect of diffusion and noise stabilizes a
finite region within the invariant manifold.

The finite nature of the  two-patch system ensures that it will
reach the absorbing state (hit the walls) as $t \to \infty$.
However, if there is an attractive region in the four dimensional
phase space, the death of the system is caused by rare events and
the typical death times  grow considerably.  For any noisy two-patch
system $Q(t)$ histograms (like those shown for a single patch in
Figure 4) were used to extract the typical decay time $\tau(D,
\Delta)$  by fitting its tail to exponential decay $exp(- t/\tau)$.
In Figure 1, $\tau(D, \Delta)$ is plotted against $D$ for different
noise amplitudes, and is shown to increase (faster than
exponentially) with $D$ ($1/D$) as $D$ approaches zero (infinity).
Evidently,  for intermediate diffusivities, an attractive manifold
appears in phase space, with a Lyapunov exponent that grows (faster
than linearly) with the diffusion constant.

\begin{figure}
  \includegraphics[width=9cm]{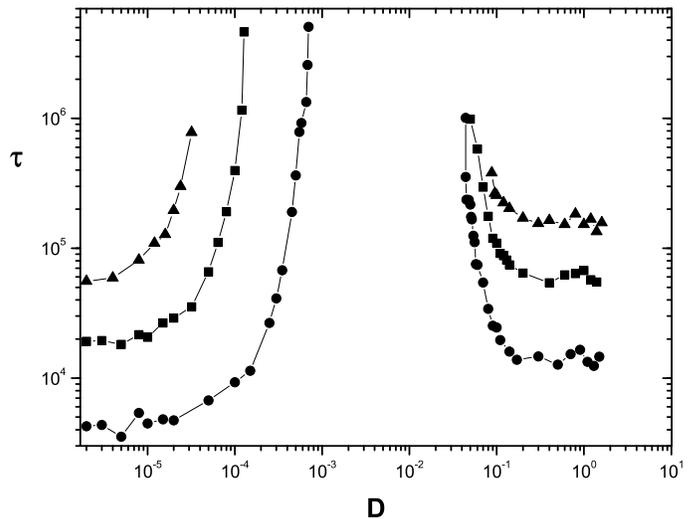}
  \caption{The typical persistence time as a function of the diffusion
rate for different levels of noise.
  The values of $\tau$ were gathered from survival probability plots (like those in Figure  3) and are displayed here
  for the two-patch system. One sees that the value of $\tau$ grows very rapidly (even faster than exponentially)
  with the migration rate for small diffusion values, and decays with $D$ for large diffusivities. Data is shown for different
  noise intensities $\Delta = 0.3$
   (triangles), $0.5$ (squares) and $1.0$ (circles).}
\end{figure}

On the other hand, at the vicinity of the homogenous fixed point,
the dynamic is similar to a single patch dynamic. The square of the
average distance from the fixed point grows linearly with time at
the beginning, with a slope that depends on the noise amplitude, as
expected for the random walk in the invariant manifold scenario. For
 "intermediate"  migration (e.g., $D=0.01$), the average
distance from the origin saturates, while the chance to find the
system at large $H$ becomes exponentially small, as illustrated in
Figure 2. In analogy with the results of the  toy model, the flow
toward the center is correlated with the desynchronization, leading
to stabilization of a soft limit cycle at finite $H$. As predicted,
while the width of the $\phi^2$ distribution depends strongly on the
noise amplitude, the oscillation amplitude is almost noise
independent.

\begin{figure}
  \includegraphics[width=8cm]{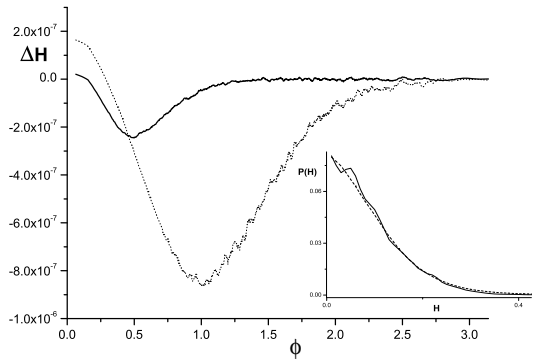}
   \caption{The average $\Delta H$ at an elementary time step (0.001 of a unit time) as a function of the angle $\phi$ between the
   patches. While a simple phase space random
  walk yields on average positive $\Delta H$, this property is shown here to hold only for small $\phi$. At larger
  angles,
  the diffusion between patches forces the system toward the center and the average $\Delta H$ becomes negative.
  Results are shown for $\Delta = 0.1$ (full line)
  and $\Delta = 1$ (dashed line). The inset shows the probability distribution function for $H$ at these two noise levels.}
\end{figure}

In conclusion, we suggest a novel solution to a long-standing
conundrum: the stabilization of a noisy unstable dynamical system on
spatial domains. The basic feature that leads to stabilization is
the dependence of  angular velocity on phase space coordinates. This
dependence allows the noise to desynchronize spatially-coupled
patches, and then migration decreases concentration gradients and
leads to stabilization of a soft limit cycle close to the homogenous
manifold.  Clearly, owing to the  phase space confinement of the
trajectories, the exact nature of the noise becomes quite
irrelevant, and thus our conclusions are also applicable to
multiplicative noise (individual based dynamics). Furthermore, the
effect of rare events - the trapping of the dynamics in the
absorbing state for any finite system - should  vanish in the
thermodynamic limit, and thus one may expect a  parametric dependent
phase transition in that limit.

\begin{acknowledgements}
 we acknowledge helpful discussions with David
Kessler, Uwe T$\ddot{\textrm{a}}$uber and  Arkady Pikovsky. This
work was supported by the Israeli Science Foundation (grant no.
281/03) and the EU 6th framework CO3 pathfinder.

\end{acknowledgements}

\appendix
\section{The noisy Lotka-Volterra model }

The Lotka-Volterra predator-prey system is a paradigmatic model for
oscillations in population dynamics \cite{lotka, volterra,murray}.
It describes the time evolution of  two interacting populations: a
prey ($b$) population that grows with a constant birth rate $\sigma$
in the absence of a predator (the energy resources consumed by the
prey are assumed to be inexhaustible), while the predator population
($a$) decays (with death rate $\mu$),  in the absence of a prey.
Upon encounter, the predator may consume the prey with  a certain
probability. Following a consumption event,  the predator population
grows and the prey population decreases. For a well-mixed
population, the corresponding partial differential equations are:
\begin{eqnarray}\label{basic}
\frac{\partial a}{\partial t} &=& - \mu a + \lambda_1 a b \\
\nonumber \frac{\partial b}{\partial t} &=& \sigma b - \lambda_2 a b
\end{eqnarray}
where $\lambda_1$ and $\lambda_2$ are the relative increase
(decrease) of the predator (prey) populations due to the interaction
between species, correspondingly.

The system admits two unstable fixed points: the absorbing state
$a=b=0$ and the state $a=0, \ \ b = \infty$. There is one marginally
stable fixed point at $\bar{a}  = \sigma/\lambda_2, \ \ \bar{b} =
\mu/\lambda_1$. Local stability analysis yields the eigenvalues $\pm
 i \sqrt{\mu \sigma}$ for the stability matrix. Moreover, even
 beyond the linear regime there is neither convergence nor
 repulsion. Using logarithmic variables $z = ln(a), \  q = ln(b)$
 eqs. (\ref{basic}) take the canonical form $\dot{z} = \partial
 H/\partial q, \ \ \dot{q} = -\partial
 H/\partial z$, where the conserved quantity $H$ (in the $ab$
 representation) is:
 \begin{equation}\label{H}
 H = \lambda_1 b + \lambda_2 a - \mu \ ln(a) - \sigma \  ln(b).
 \end{equation}
The phase space, thus, is segregated into a collection of nested
one-dimensional trajectories, where each one is characterized by a
different value of  $H$, as illustrated in Figure 3.  Given a line
connecting the fixed point to one of the "walls" (e.g., the dashed
line in the phase space portrait, Figure 3), $H$ is a monotonic
function on that line, taking its minimum $H_{min}$ at the
marginally stable fixed point (center) and diverging on the wall.
Without loss of generality, we employ hereon the symmetric
parameters $\mu = \sigma = \lambda_1 = \lambda_2 =1$. The
corresponding phase space, along with the dependence  of $H$ on  the
distance from the center and a plot of the oscillation period vs.
$H$, are represented in Figure 3).

\begin{figure}
  \includegraphics[width=8cm]{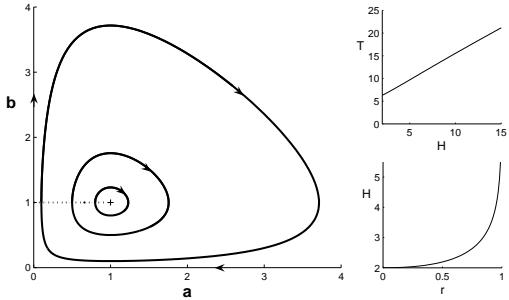}
  \caption{The Lotka-Volterra phase space (left panel) admits a
  marginally stable fixed point surrounded by close trajectories (three of these
  are
  plotted). Each trajectory corresponds to single $H$ defined in Eq.
  (\ref{H}), where $H$ increases monotonically  along the (dashed) line
  connecting the center with the $a=0$ wall, as shown in the lower
  right panel. In the upper right  panel, the period of a cycle $T$ is
  plotted against $H$, and is shown to increase almost linearly from
  its initial value $T = 2 \pi / \sqrt{\mu \sigma}$ close to the
  center.}
  \label{fig3}
\end{figure}

Given the integrability of that system, the effect of noise is quite
trivial: if $a$ and $b$ randomly fluctuate in time (e.g., by adding
or subtracting small amounts of population during each time step),
the system wanders between trajectories, thus performing some sort
of random walk in  $H$  with "repelling boundary conditions" at
$H_{min}$ and "absorbing boundary conditions" on the walls (as
negative densities are meaningless, the "death" of the system is
declared when the trajectory hits the zero population state for one
of the species). This result was emphasized by  Gillespie
\cite{gillaspee} for the important case where intrinsic stochastic
fluctuations are induced by the discrete character of the reactants.
In that case, the noise is multiplicative (proportional to the
number of particles), and the system flows away from the center and
eventually hits one of the absorbing states at $0,0$ or $0,\infty$.
The corresponding situation for a single patch Lotka-Volterra system
with additive noise is demonstrated in Figure 4, where the survival
probability $Q(t)$ (the probability that a trajectory does not hit
the absorbing walls within  time $t$) is shown for different noise
amplitudes.

\begin{figure}
  \includegraphics[width=7cm]{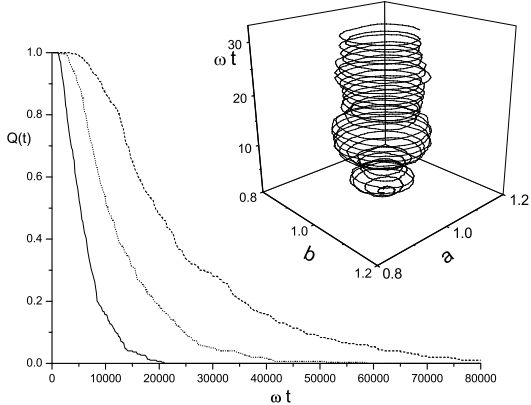}
 \caption{The survival probability $Q(t)$ is plotted versus time for
a single patch noisy LV
  system. Eqs. (\ref{basic}) (with the symmetric parameters) were integrated numerically (Euler integration
  with time step $0.001$), where the initial conditions are at the fixed point $a=b=1$. At each time step, a small random number
  $\eta(t) \Delta t$ was added to each population density, where $\eta(t) \in [-\Delta,\Delta]$. A typical phase space trajectory, for
  $\Delta = 0.5$, is shown in the inset. The system "dies" when the trajectory hits the walls $a=0$ or $b=0$. Using 300 different noise histories, the
  survival probability is shown here for $\Delta = 0.5$ (full line), $\Delta = 0.3$ (dotted line) and $\Delta = 0.25$ (dashed line).
  Clearly, the survival probability decays exponentially at long
  times, $Q(t) \sim exp(-t/\tau)$. As expected for a random walk
  with absorbing boundary conditions, $1/ \tau$ scales with
  $\Delta^2$.}
  \label{fig2}
\end{figure}

\section{Coupled oscillators}

The Lotka-Volterra system (Appendix A) is somewhat complicated,
since the angular velocity depends not only on $H$,  but also on the
location along a trajectory. In order to simplify the discussion,
let us introduce a toy model that imitates the main features  of the
real systems. Although that model does not allow for an absorbing
state, we believe that it captures the basic mechanism for
stabilization of spatially extended systems in the presence of
noise.

The toy model deals with the phase space behavior of diffusively
coupled oscillators, where the angular frequency depends on the
radius of oscillations. With additive noise, the Langevin equations
take the form:

\begin{eqnarray}\label{oscillators}
\frac{\partial x_1}{\partial t} &=& \omega(x_1,y_1) y_1 + D_1
(x_2-x_1) + \eta_1(t) \nonumber
\\ \nonumber \frac{\partial x_2}{\partial t} &=& \omega(x_2,y_2) y_2+
D_1 (x_1-x_2)+ \eta_2(t)  \\ \frac{\partial y_1}{\partial t} &=&
-\omega(x_1,y_1) x_1 + D_2 (y_2 - y_1)+ \eta_3(t) \\ \nonumber
\nonumber \frac{\partial y_2}{\partial t} &=& -\omega(x_2,y_2) x_2 +
D_2 (y_1 - y_2) + \eta_4(t).
\end{eqnarray}
where all the $\eta$-s are taken from the same distribution.  If the
angular frequency is location independent, $\omega(x,y) = \omega_0$,
the problem is reduced to coupled \emph{harmonic} oscillators, a
diagonalizable linear problem that admits two purely imaginary
eigenvalues in the invariant, homogenous manifold. With the addition
of noise, the random walk on that manifold is independent of the
motion in the fast manifold, such that the radius of oscillation
diverges with the square root of time. As there are no "absorbing
walls" here, the oscillation amplitude will grow indefinitely in the
presence of noise.

Now let us define the oscillation radius for each patch, $r_i =
\sqrt{x_i^2 + y_i^2}$ for $i=1,2$, and assume that the angular
frequency depends only on that radius and is $\theta$-independent
[$\theta_i \equiv arctg(y_i/x_i)$]. With that, the total phase $\Phi
= \theta_1 + \theta_2$ decouples and the 3-dimensional phase space
motion is dictated by the equations (we assume $D_1 = D_2 = D$ and
define $\phi = \theta_1 - \theta_2$):
\begin{eqnarray}\label{3d}
 \dot{r_1} &=& -D(r_1 - r_2 \cos(\phi)) + \tilde{\eta}_{1}(t)
\\  \nonumber  \dot{r_2}  &=& -D(r_2 - r_1 \cos(\phi)) + \tilde{\eta}_{2}(t)\\
 \dot{\phi} = -&D&
 \left( \frac{r_1^2+r_2^2}{r_1 r_2} \right) sin \phi - \left[ \omega(r_1) -
\omega(r_2) \right]+  \left( \frac{\tilde{\eta}_{3}}{r_1} -
\frac{\tilde{\eta}_{4}}{r_2} \right) \nonumber.
\end{eqnarray}
As before, all the $\tilde{\eta}$-s are taken from the same
distribution. In the harmonic limit, $\omega(r_1) = \omega(r_2) =
\omega_0$, the only noisy term for $\phi$ vanishes at large  $r$-s,
and thus $\phi$ approaches zero. Defining now the new coordinates,
$R \equiv r_1+r_2$, $r \equiv  r_1 -r_2$,
\begin{eqnarray} \label{Rr}
\dot{R} = -2D sin^2(\phi/2)R + \eta  \\ \dot{r} = -2Dcos^2(\phi/2)r
+ \eta,
\end{eqnarray}
one notices that at the limit $\phi = 0$ the "restoring force" in
the $R$ direction vanishes  and the phase space admits an attracting
1d-manifold $\phi = r = 0$ (the invariant manifold). The noise,
thus, induces random walk on that $R$ manifold, and since the walker
cannot cross the origin, its displacement grows like $\sqrt{t}$.

\begin{figure}
  \includegraphics[width=8cm]{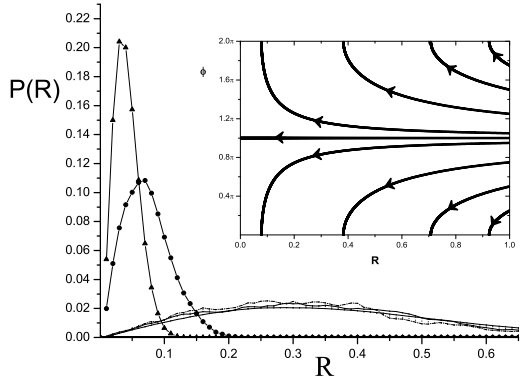}
  \caption{Histograms showing the probability to be at a distance $R$ from the origin as a function of $R$, for
  two coupled noisy oscillators, where $\omega = 1+ \alpha r$  with $D=0.01$ and various values of noise strength $\Delta$ and
  angular velocity gradient $\alpha$. As expected, the phase space confinement is proportional to $\alpha$, from $\alpha = 1$ (triangles)
 to $\alpha = 0.5$ (circles) to $\alpha = 0.1$ (solid line), all for the same level of noise $\Delta = 0.1$. On the other hand,
 as predicted by the linear analysis close
 to the invariant manifold, the confinement is noise independent, and the three solid lines corresponding to
 different levels of noise ($\Delta = 0.1,0.5,1$ with the same $\alpha=0.1$ almost coincide. The inset shows the flow lines
 on the $r=0$ plane. The invariant manifold $\phi =0$ is stable, but there is a "maximally desynchronized" unstable orbit
 converging to the center at $\phi = \pi$. If the expectation value of $\phi^2$ deviates from zero, there is an effective restoring force
 toward the center, and the noise-induced random walk on the $\phi =0$ manifold is bounded. }
  \label{fig4}
\end{figure}

In the generic case, however, where $\omega$ depends on $r$,  the
system behaves quite differently: if $r_1 \neq r_2$, the two patches
oscillate with different radial distance and \emph{desynchronize},
i.e.,  $\langle \phi^2 \rangle$ acquires, on average, some finite
value. Migration, then, acts as a restoring force for an overdamped
harmonic oscillator [Eq. (\ref{Rr})] and stabilizes the oscillations
(if diffusion stabilizes the unstable fixed point this (noise
independent) phenomenon is known as "oscillation death"
\cite{bareli}).

Following Eq. (\ref{Rr}), close to the homogenous manifold (large
diffusion, small noise limit), $r^2$ typical fluctuation around zero
is of order $\Delta^2/D$. The $\omega(r_1) - \omega(r_2)$ term
induces finite noise in the $\phi$ equation of motion, and thus
$\langle \phi^2 \rangle \sim (\omega^{'}(r))^2 \Delta^2 / D^3$. This
desynchronization, in turn, leads to  the appearance of a finite
restoring force on the invariant manifold $R$, as $\langle \phi^2
\rangle \neq 0$, thus (far from the origin) one finds $\langle R^2
\rangle \sim D^2 /\omega^{'}(r))^2 $. The small $D$ instability
(decoupled patches) manifest itself in the divergence of $r^2$ as $D
\to 0$.   It should be noted that, since both the restoring force
and the noise in the invariant manifold are proportional to
$\Delta^2$, the expected $R$ distribution has to be \emph{noise
independent} at that limit, as demonstrated in Figures 5 and 2. All
in all, if $\omega$ is radius dependent, the coupled oscillators'
system stabilizes at some finite radius from the origin, giving rise
to a soft "limit cycle" in the 4-dimensional phase space, as
indicated in Figure 5. The above considerations are valid only close
to the homogenous manifold. For stronger noise, although the
qualitative features of the system are the same,  it was shown
recently that \cite{mckane} the effect of noise may shift the
"intrinsic" frequency of the system, leading to some shift of the
intrinsic frequency.

Eqs. (\ref{oscillators}) may be generalized to include the case of
an unstable focus (in order to imitate the dynamics of the
Nicholson-Bailey \cite{bailey}  map) by adding a diagonal repulsive
term to any variable (e.g., $ \dot{x_1} =  \omega(x_1,y_1) y_1 + D_1
(x_2-x_1) + \eta_1(t) + \epsilon x_1$, where $\epsilon$ measures the
"strength" of the repulsion). The same analysis shows that, for
small noise and small repulsion, a noise-induced transition will
occur at $\epsilon \sim (\omega'^2 \ \Delta^2 /D^2)$. If the noise
is small enough, the desynchronization is weak and cannot stabilize
a soft limit cycle, and thus the system is, for real populations,
extinction-prone. Strong noise, conversely, stabilizes the system
and ensures species conservation.

\end{document}